\documentclass[twocolumn,amsmath,amssymb,superscriptaddress]{revtex4}

\usepackage{graphicx}
\usepackage{dcolumn}
\usepackage{bm}

\begin{document}

PRL-Viewpoints: Physics \textbf{2}, 108 (2009).

http://physics.aps.org/articles/v2/108

\preprint{APS/123-QED}

\begin{abstract}

The energy-momentum relationship of electrons on the surface of an ideal topological insulator forms a cone - a Dirac cone, which, when warped (no longer described by the Dirac equation), can lead to unusual phenomena such as enhanced electronic interference around defects and a magnetically ordered broken symmetry surface. A detailed spin-texture and hexagonal warping maps on Bi$_2$Te$_3$ are presented here.

\end{abstract}

\title{Spin-textures, Berry's phase and Quasiparticle Interference in Bi$_2$Te$_3$: A Topological Insulator with Warped Surface States}

\author{M.Z. Hasan}
\affiliation{Joseph Henry Laboratories : Department of Physics, Princeton University, Princeton, NJ 08544}
\affiliation{Princeton Center for Complex Materials, Princeton University, Princeton, NJ 08544\*}
\email{mzhasan@Princeton.edu}
\author{H. Lin}
\affiliation{Department of Physics, Northeastern University, Boston, MA 02115}
\author{A. Bansil}
\affiliation{Department of Physics, Northeastern University, Boston, MA 02115}

\date{\today}

\maketitle

\begin{figure*}
\includegraphics[width=0.5\textwidth,angle=-90]{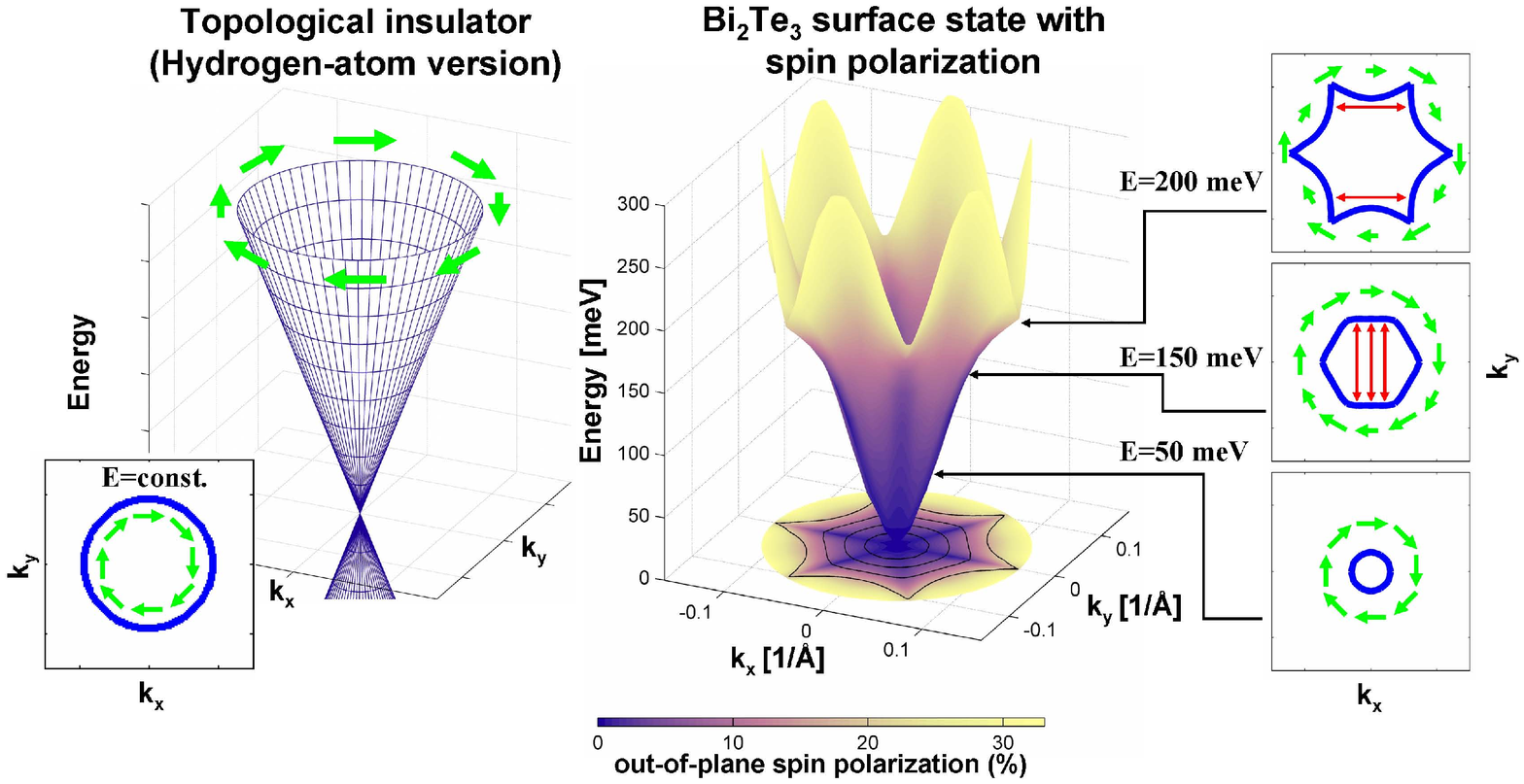}
\caption{\label{fig:fermi} \textbf{Spin-textures, Berry's phase and Quasiparticle Interference in Bi$_2$Te$_3$}: (Left) In the simplest ("hydrogen-atom") case, the energy-momentum relationship of the surface states in a topological insulator takes the form of a Dirac cone. The constant energy surfaces are then circles of different radii. (Right) Results of a first-principles computation of the spin texture of the surface states in Bi$_2$Te$_3$ that we have carried out, showing in quantitative detail how the Dirac cone is warped due to the effect of the crystal potential. The actual shape of the Fermi surface is determined by the natural chemical potential and can be a hexagon or a snowflake. As a result of deviations from the ideal Dirac cone dispersion, a nonzero out-of-the plane spin polarization develops and the conventional nesting channels (red arrows) open up possibilities for magnetic order on the surface.}
\end{figure*}

Unlike common ordered states of matter such as the crystalline solids and magnets characterized by some broken symmetry, the topological states of quantum matter are characterized by boundary properties that are highly robust or stable in the presence of disorder, fluctuations, and perturbations \cite{1, 2}. The first example of a two-dimensional topological insulator was provided by the quantum Hall effect in an electron gas \cite{1}, which has no distinct three-dimensional topological generalization \cite{3, 4, 5, 6}. Recent theoretical and experimental work indicates that strong spin-orbit coupling in insulators can produce a three-dimensional topological insulator, which is a completely new state of matter \cite{3, 4, 5, 6, 7, 8, 9, 10, 11, 12, 13}. This state is characterized by a surface Berry's phase, realized via an odd number of Dirac cones protected against disorder such as alloying \cite{3, 4, 5, 6}. This is in sharp contrast to the case of graphene, which possesses an even number of Dirac cones.

The thermoelectric Bi-Sb alloys discovered to be topological last year provided the first example of a truly three-dimensional topological insulator \cite{7, 8, 9}. The experimental discovery of another thermoelectric family followed the Bi$_2$X$_3$ (X=Se,Te) series - harboring the same topological insulator states with a simpler surface spectrum \cite{10, 11, 12, 13}. The large band gap (0.3 eV) topological insulator Bi$_2$Se$_3$ features a nearly ideal Dirac cone \cite{10, 12}, but in the smaller band gap (0.15 eV) material Bi$_2$Te$_3$, with trigonal crystal potential, the cone is warped \cite{11, 12}. Even though the cone is deformed with hexagonal symmetry, the Berry's phase quantifying the topological invariant remains unchanged, consistent with its topological order \cite{12, 13}. Now, in an article appearing in the current issue of Physical Review Letters \cite{14}, Liang Fu of Harvard University in the US presents a perturbation calculation, which yields a hexagonal modulation of the surface states analogous to the existence of a nonlinear term in the spin-orbit interaction in the bulk crystal. (See Fig. 1 for an illustration using our results of a full-potential, self-consistent computation in Bi$_2$Te$_3$.) This leads to a number of remarkable predictions concerning the unusual physical properties induced by the warping of the surface states, which are distinct from the simpler "hydrogen-atom" version of the topological insulator discovered previously \cite{10}.

The spin and linear momentum of an electron on the surface of a topological insulator are locked-in one-to-one, as revealed in recent spin-resolved photoemission measurements \cite{8, 9, 12, 13}: the electron is forbidden from moving backwards or taking a "U-turn" an effect that hinders its localization \cite{15}. This can be realized in a real material if the spins of electrons lying on the Dirac cone, but away from the apex (the Dirac node), rotate around in a circle at the Fermi level. Spins arranged in this way lead to a geometrical quantum phase known as Berry's phase with a value of $\pi$ \cite{8, 12}. The work of Fu effectively shows that in the presence of a hexagonally deformed cone, the spins must acquire finite out-of-the-plane components to conserve the net value of Berry's phase, thus preserving the bulk topological invariant. The resulting finite value of the out-of-the-plane component opens up new possibilities for observing spectacular quantum effects.

Fascinating classes of new particles such as a Majorana fermion, or a magnetic monopole image, or a massive spin-textured Dirac particle can live on the surface of a topological insulator if a finite gap can be induced in the surface band with the chemical potential near the Dirac node \cite{3, 4, 5, 6, 16}. One approach is to create an interface between the topological insulator and a more conventional material such as a magnet or a superconductor, which can induce an energy gap in the Dirac bands. Alternatively, a magnetic field can be applied perpendicular to the sample surface. Until now it had been thought that a parallel magnetic field would neither open a gap nor lead to a surface Hall effect. However, when the surface states are warped significantly in a truly bulk insulating sample, as the work of Fu shows, a parallel magnetic field would open up a gap in the surface spectrum, enabling quantum Hall experiments to be carried out using relatively modest magnetic fields (less than 7 tesla) accessible in a typical laboratory setting. This is an exciting prospect since such experiments have so far not been possible because very high magnetic fields are needed for opening a gap.

In a truly bulk insulating sample, where the chemical potential is such that the hexagonal warping of the Fermi surface survives, one would expect to see a rare realization of the quantum Hall effect on the surface without the presence of Landau levels. In this way, one could observe a half-integer quantum of conductance per surface, which has not been possible so far in any known two-dimensional electron gas, including graphene, and thus allow experimental tests of currently debated scenarios for unusual antilocalization-to-localization transitions on a topological insulator \cite{5}. Presently realized Bi$_2$Te$_3$ is quite bulk-metallic with small resistivity \cite{11, 12}, whereas spin-orbit insulators with high bulk resistivity values coupled with surface carrier control are needed to realize a functional topological insulator with which many more interesting experiments could be possible. Recent progress in making highly resistive Bi$_2$Se$_3$ \cite{12, 17} coupled with surface carrier control \cite{18} heralds the potential dawn of a topological revolution. Similar improvements in sample resistivity are also called for in Bi$_2$Te$_3$.

The hexagonal warping of the surface cone and the resulting deviations from a circular Fermi surface provide a natural explanation for the enhanced magnitude of surface electronic interference effects recently reported around crystalline defects in Bi$_2$Te$_3$ \cite{19, 20}. For a circular Fermi surface (Fig. 1, left), the electronic states at the Fermi surface cannot backscatter since in a fully spin-polarized band \cite{12, 13} a consequence of time-reversal symmetry is that there are no available states into which electrons can scatter elastically. As a result, interference effects around the defects are small. With hexagonal warping, however, new electronic scattering channels open up, connected by various simple nesting vectors on the Fermi surface (see Fig. 1), and consistent with recent scanning tunneling measurements, the scattering due to defects becomes enhanced \cite{19, 20}.

The ideal topological insulators displaying a perfect Dirac cone dispersion cannot support the formation of charge- or spin-density waves. However, in the presence of hexagonal or other forms of deformations of the surface band, such ubiquitous ordering instabilities are no longer forbidden, but rather present a competing ground state, which can circumvent the topological protection. Since electrons with opposite momenta possess opposite spins, a spin-density-wave state is favored if the Fermi surface has parallel segments that can be nested, as is seen to be the case for hexagonal deformation (see Fig 1, right). As one goes around the hexagonal Fermi surface, electrons will experience rapid changes in the out-of-the-plane spin component ($\sigma_z$) to maintain their overall Berry's phase value at $\pi$ \cite{8, 12}, making the resulting spin-density wave appear quite exotic. A stripelike one-dimensional order on the topological surface may also be energetically possible.

Even more complex spin-density-wave orderings could be supported by other topological insulators that presently lack a tractable theoretical model. For example, Bi$_{1-x}$Sb$_x$ alloys display a central hexagonally deformed Dirac cone and additional surface states of complex character \cite{8}. Despite the complexity of its five surface bands and hexagonal warping, the Bi$_{1-x}$Sb$_x$ alloy topological insulator precisely exhibits $\pi$ Berry's phase and antilocalization just like the "hydrogen-atom" single-surface-Dirac-cone systems \cite{8, 10, 12}. This is a remarkable manifestation and demonstration of a deep and profound principle of physics - topological invariance. Thus the power and beauty of the topological order and invariance in nature are rather firmly established by studying systems that have complex "geometries," i.e., complex and warped surface states as in the Bi$_{1-x}$Sb$_x$ alloy series or Bi$_2$Te$_3$.

The most remarkable property of topological surface states is that they are expected to be perfect metals in the sense that they are robust and stable against most perturbations and instabilities towards conventional ordered states of matter. The warping of the surface states, even though it opens up interesting possibilities for studying this unusual state of matter in detail, ironically also opens up competing channels which provide a route for departure from the much sought after paradigm of perfect topological stability of the spin-textured metal \cite{3, 4}. Most topological insulators lying hidden in nature are likely to harbor complex and warped surface states as in Bi$_{1-x}$Sb$_x$ and Bi$_2$Te$_3$.

Interestingly, even though the first three-dimensional topological insulators and their spin-textures carrying nontrivial Berry's phases were discovered only recently, we already have new materials emerging with physical properties lying in a regime where quantum effects associated with competing orders such as the spin-density waves can be foreseen. Parallel progress in theoretical modeling, hand-in-hand with experimental advances in spin-sensitive measurement methods, will likely be essential for unraveling and eventually exploiting the exotic spin textures \cite{21, 22} and other beautiful properties of topological quantum matter \cite{23}.

\begin{figure*}
\includegraphics[width=0.53\textwidth,angle=-90]{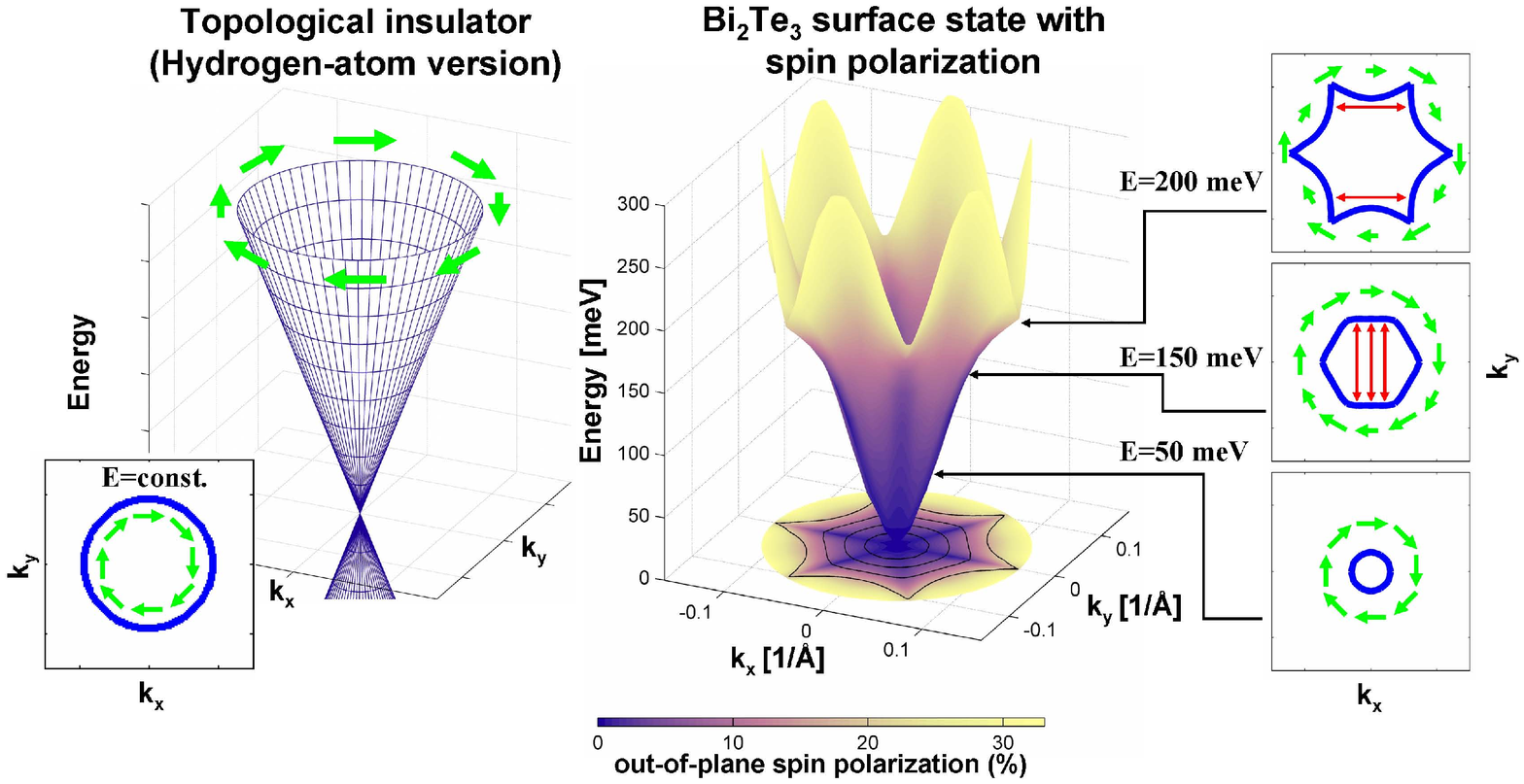}
\caption{\label{fig:fermi} \textbf{Spin-textures, Berry's phase and Topological Invariance in Bi$_2$Te$_3$}:(Left) In the simplest ("hydrogen-atom") case, the energy-momentum relationship of the surface states in a topological insulator takes the form of a Dirac cone. The constant energy surfaces are then circles of different radii. (Right) Results of a first-principles computation of the spin texture of the surface states in Bi$_2$Te$_3$ that we have carried out, showing in quantitative detail how the Dirac cone is warped due to the effect of the crystal potential. The actual shape of the Fermi surface is determined by the natural chemical potential and can be a hexagon or a snowflake. As a result of deviations from the ideal Dirac cone dispersion, a nonzero out-of-the plane spin polarization develops and the conventional nesting channels (red arrows) open up possibilities for magnetic order on the surface. [M.Z. Hasan, H. Lin, A. Bansil, Physics \textbf{2}, 108 (2009)]}
\end{figure*}

\end{document}